\documentclass[aps,prd,preprint,showpacs,showkeys]{revtex4}
\usepackage{graphics} 
\newcommand{\V}{\mathcal{V}}
\newcommand{\e}{\mathrm{e}}
\newcommand{\ep}{\epsilon}
\newcommand{\vev}[1]{\left\langle #1 \right\rangle}
\begin{document}
\preprint{KOBE-TH-02-06}
\title{Bootstrapping Perturbative Perfect Actions}
\author{Hidenori Sonoda}
\affiliation{Physics Department, Kobe University, Kobe 657-8501,
Japan}
\email[E-mail address: ]{sonoda@phys.sci.kobe-u.ac.jp}
\date{25 December 2002}
\begin{abstract}
We study the exact renormalization group of the four dimensional
$\phi^4$ theory perturbatively.  We reformulate the differential
renormalization group equations as integral equations that define the
continuum limit of the theory directly with no need for a bare theory.
We show how the self-consistency of the integral equations leads to
the determination of the interaction vertices in the continuum limit.
The inductive proof of the existence of a solution to the integral
equations amounts to a proof of perturbative renormalizability, and it
consists of nothing more than counting the scale dimensions of the
interaction vertices.  Universality is discussed within a context of
the exact renormalization group.
\end{abstract}
\pacs{05.10.Cc, 11.10.Gh, 11.10.Hi}
\keywords{exact renormalization group, perfect action}
\maketitle

\section{Introduction}

Renormalization theory has more than fifty years of history starting
from the studies of ultraviolet divergences in QED.  Originally
thought of as a cook book recipe for obtaining finite results free of
ultraviolet divergences, the idea of renormalization took a long time
for its full acceptance until its physical meaning was clarified, and
the relation to universality in critical phenomena was understood.  In
his series of lectures \cite{wk74} Wilson explains how to construct
the continuum limit of a quantum field theory by taking a classical
statistical model to the limit of criticality.  The so-called scaling
functions are what particle physicists call renormalized Green
functions from which scattering cross sections of elementary particles
can be computed.

One thread of development, lattice simulation of field theory, was
started immediately after Wilson's work on renormalization, and the
field thrives to this day.  Another development started somewhat later
when Polchinski applied Wilson's exact renormalization group (ERG)
equation to the study of perturbative renormalization.\cite{pol84}

The central idea in Wilson's renormalization theory is the theory
space that consists of all possible theories with a fixed cutoff
scheme.  Flows of the renormalization group are generated in the
theory space under the rescaling of distance.  To keep track of the
renormalization group flows exactly, the theory space must contain an
infinite number of dimensions allowing for all possible interaction
vertices.  But only a finite dimensional subspace, denoted as
$S(\infty)$ in sect.~12 of Ref.~\onlinecite{wk74}, are of fundamental
importance.  This is the space of flows originating from an
ultraviolet fixed point.  It is parameterized by a finite number of
parameters, called relevant parameters.  Any theory in this subspace
can be traced backward along a renormalization group flow to the fixed
point, and the theory gives a continuum limit.  In more recent
literature, the theories in $S(\infty)$ are called (quantum) perfect
actions, implying that they contain the physics of continuous space
despite the use of a finite momentum cutoff.  For a review on perfect
actions, see Ref.~\onlinecite{has98} and references therein.

In Ref.~\onlinecite{pol84} Polchinski rendered Wilson's exact
differential renormalization group equations to a form more manageable
for perturbative studies.  Polchinski used his form of equations to
obtain a quantitative estimate for how the flows from bare theories
approach $S(\infty)$.  Strictly speaking the theory considered,
$\phi^4$, has no ultraviolet fixed point, but $S(\infty)$ exists
perturbatively, and the distance between the flows and $S(\infty)$ has
been shown to behave as $\e^{-2t}$, where $t$ is the logarithmic
momentum scale so that the physical cutoff momentum is $\e^t$ times
the physical renormalization scale.\footnote{At each order of
perturbation theory, the approach is modified by a finite integral
power of $t$.}  Polchinski's work brought Wilson's physical insight
into renormalization to the perturbative renormalization theory which
had been mostly diagrammatic and calculational.

The purpose of the present work is to simplify the perturbative study
of the exact renormalization group (ERG) even further by reformulating
the ERG differential equations as integral equations that define the
continuum limit $S(\infty)$ directly.  As is well known, an integral
equation is nothing more than a differential equation together with an
initial (or asymptotic) condition, but the rewriting brings a great
advantage in this case.  The advantage is that the integral equations
incorporate ``renormalizability'' of the theory manifestly.  If the
equations have a solution, the theory is renormalizable automatically.
The issue is not the ultraviolet finiteness of the theory, but it is
the existence of a solution.

The existence of a solution to the integral equations is proved using
perturbation theory.  A recursive solution of the integral equations
is what we call perturbation theory.  The integral equations are
self-contained, and can determine themselves.  Using the word
``bootstrap'' as a mnemonic for the self-determining nature, we can
say that the integral equations bootstrap themselves.

A careful examination of the original work of Polchinski has been made
in Ref.~\onlinecite{bt94} where the issues of perturbative analyticity,
unitarity, and causality have also been studied.  The emphasis of the
present paper is on the new formulation of the ERG in terms of
integral equations and on the new insights given by the formulation,
and we do not aim at the rigor exemplified in Ref.~\onlinecite{bt94}.  Our
``proof'' in section \ref{solution} is a ``physicist's proof'' which
should be acceptable to almost any physicist.

The present paper is organized as follows.  In sect.~\ref{ERG} we
review the perturbative treatment of Wilson's ERG by Polchinski.  In
sect.~\ref{integral} we introduce the reformulation of the ERG as
integral equations.  In sect.~\ref{solution} we prove inductively that
the integral equations have a solution.  In sect.~\ref{univ} we
discuss universality in the context of the integral ERG equations.
Finally in sect.~\ref{conclusion} we conclude the paper with comments
for further developments of the integral equation approach.

\section{\label{ERG}Exact renormalization group equations}

A large amount of literature is available on the exact renormalization
group (see Ref.~\onlinecite{bb01} and references therein), and the main
purpose of this section is to set the notation for the rest of the
paper.  Please note that unlike what is common in the field theory
literature, a rescaling is done after each step of renormalization to
keep the momentum cutoff constant.

\subsection{Brief review of Polchinski's rendition of
Wilson's differential ERG equation}

We consider a $\mathbf{Z}_2$ invariant scalar field theory in four
dimensional euclidean space.  The propagator of the scalar field
$\phi$ is given by
\begin{equation}
\frac{K(p)}{p^2 + m^2}
\end{equation}
where the momentum cutoff function $K(p)$ is a smooth scalar function
that is monotonically decreasing in $p^2$ and has the property
\begin{equation}
K(p) = \cases{1 &for $p^2 < 1$\cr
0 &for $p^2 > 2^2$\cr}
\end{equation}
The cutoff function $K(p)$ is fixed once and for all for all the
theories in the theory space.

Each point in the theory space corresponds to an interaction action
given in the following form:
\begin{equation}
- S_\mathrm{int} [\phi] = \sum_{n=1}^\infty \frac{1}{(2n)!}
\int_{p_1,\cdots,p_{2n-1}} \phi (p_1) \cdots \phi (p_{2n}) \, \V_{2n}
(p_1,\cdots,p_{2n}) \label{Sint}
\end{equation}
where $p_{2n} \equiv - (p_1 + \cdots + p_{2n-1})$, and the momentum
integrals are taken only over the $2n-1$ independent momenta.  The
notation 
\begin{equation}
\int_p f(p) \equiv \int \frac{d^4 p}{(2\pi)^4} f(p)
\end{equation}
is used for the momentum integral.  We will call $\V_{2n}$ an
interaction vertex from now on.  $\phi (p)$ is the Fourier transform
of the scalar field in the momentum space.  Once the interaction
action $S_\mathrm{int}$ is given, the generating functional of the
Green functions is obtained as
\begin{equation}
Z[J] = \exp \left( \frac{1}{2} \int_p \frac{K(p)}{p^2+m^2}
\frac{\delta}{\delta \phi (p)} \frac{\delta}{\delta \phi (-p)} \right)
\e^{- S_\mathrm{int} [\phi] + \int_p J(-p) \phi (p)} \Bigg|_{\phi = 0}
\end{equation}
and the $2n$-point Green function is given by
\begin{eqnarray}
&&\vev{\phi (p_1) \cdots \phi (p_{2n-1}) \phi}_{m^2, \V} \nonumber\\
&=& \exp \left(
\frac{1}{2} \int_p \frac{K(p)}{p^2+m^2} \frac{\delta}{\delta \phi (p)}
\frac{\delta}{\delta \phi (-p)} \right) \, \phi(p_1) \cdots \phi
(p_{2n-1}) \phi \,\e^{- S_\mathrm{int} [\phi]} \Bigg|_{\phi = 0}
\end{eqnarray}

Since the cutoff function $K$ is fixed, a theory is specified by the
choice of the squared mass $m^2$ and the interaction action
$S_\mathrm{int}$.  The latter is characterized by an infinite number
of interaction vertices $\{\V_{2n}\}$, and the theory space is
infinite dimensional.

The ERG transformation by scale $\e^{\Delta t}$, where $\Delta t$ is
an infinitesimal positive constant, is defined so that the momentum
$p$ corresponds to the higher momentum $p \e^{\Delta t}$ under the
transformation.  In the case of free theory, all the vertices
$\{\V_{2n}\}$ vanish, and only the squared mass scales as
\begin{equation}
m^2 \longrightarrow m^2 \e^{2 \Delta t}
\end{equation}
under the renormalization.  In the presence of interactions, we must
transform the vertices $\{\V_{2n}\}$ to $\{\V_{2n} + \Delta \V_{2n}\}$
so that the Green functions are related by
\begin{eqnarray}
&&\vev{ \phi (p_1) \cdots \phi (p_{2n-1}) \phi }_{m^2, \V} \nonumber\\
&=& \e^{(4n - y_{2n})\Delta t} \vev{ \phi
(p_1 \e^{\Delta t}) \cdots \phi (p_{2n-1} \e^{\Delta t}) \phi }_{m^2
\e^{2 \Delta t}, \V + \Delta \V} \label{scaling}
\end{eqnarray}
where we define
\begin{equation}
y_{2n} \equiv 4 - 2n \label{scaledim}
\end{equation}

The infinitesimal change of the interaction action $\Delta
S_\mathrm{int}[\phi]$ corresponding to the infinitesimal change of
the interaction vertices $\{\Delta \V_{2n}\}$ was given by Wilson in
Ref.~\onlinecite{wk74}.  In this paper we will consider the particular
form given by Polchinski in Ref.~\onlinecite{pol84}:
\begin{equation}
- \Delta S_\mathrm{int} = \Delta t \cdot \frac{1}{2} \int_p \frac{- 2
p^2 \frac{d K(p)}{dp^2}}{p^2+m^2} \, \left\{ \frac{\delta
S_\mathrm{int}}{\delta \phi (-p)} \frac{\delta S_\mathrm{int}}{\delta
\phi (p)} - \frac{\delta^2 S_\mathrm{int}}{\delta \phi (p) \delta \phi
(-p)} \right\} \label{polchinski}
\end{equation}
(This is exactly the same as Eq.~(18) of Ref.~\onlinecite{pol84}
rewritten in our notation.)  We can obtain the ERG equations for the
individual vertices $\{\V_{2n}\}$ by substituting the perturbative
expansion (\ref{Sint}) into the above.

Let us introduce a logarithmic scale parameter $t$ to define a
one-parameter family of vertices $\{\V_{2n} (t)\}$.  At $t$, the
squared mass is given by $m^2 \e^{2t}$.  Polchinski's equation
(\ref{polchinski}) implies the following differential equations for
the vertices:
\begin{eqnarray}
&&\frac{d}{dt} \left( \e^{- y_{2n}t} \V_{2n} (t; p_1 \e^{t},\cdots,
p_{2n} \e^{t}) \right)\nonumber\\ &=&
\sum_{k=0}^{\left[\frac{n-1}{2}\right]}
\sum_{\text{partitions:} \atop I+J=\{2n\}} \e^{- y_{2(k+1)} t}
\V_{2(k+1)} (t; p_{I_1} \e^t,\cdots, p_{I_{2k+1}} \e^t, -
(p_{I_1}+\cdots+p_{I_{2k+1}}) \e^t )\nonumber\\ &&\qquad\qquad\qquad
\times\, \frac{\Delta ((p_{I_1} + \cdots + p_{I_{2k+1}})^2
\e^{2t})}{(p_{I_1} + \cdots + p_{I_{2k+1}})^2 + m^2} \nonumber\\ &&
\qquad\qquad\qquad \times \,\e^{- y_{2(n-k)} t}
\V_{2(n-k)} (t; p_{J_1} \e^t , \cdots, p_{J_{2(n-k)-1}} \e^t,
(p_{I_1}+\cdots+p_{I_{2k+1}}) \e^t ) \nonumber\\ && \, + \frac{1}{2}
\int_q \,\frac{\Delta (q \e^t)}{q^2 + m^2} \,\e^{- y_{2(n+1)} t}
\V_{2(n+1)} (t; q \e^t, - q \e^t, p_1 \e^t, \cdots, p_{2n} \e^t)
\label{diffeq}
\end{eqnarray}
where the sum over partitions is the sum over all possible ways of
splitting $p_1, \cdots, p_{2n}$ into two groups, and $I$ and $J$ stand
for the groups of $2k+1$, $2(n-k)-1$ elements, respectively.  Later we
will introduce a short hand notation $p_I$ to mean either the list of
$p_{I_1}, \cdots, p_{I_{2k+1}}$ or the sum $p_{I_1} + \cdots +
p_{I_{2k+1}}$.  The same goes for $p_J$.  The function $\Delta (p)$ is
defined by
\begin{equation}
\Delta (p) \equiv - 2 p^2 \frac{d}{dp^2} K(p)
\end{equation}
The Gauss symbol $[(n-1)/2]$ denotes the largest integer less than or
equal to $(n-1)/2$.

Two comments are in order:
\begin{enumerate}
\item Given the requirement (\ref{scaling}), the ERG transformation is
not uniquely determined due to a potential change of fields, and it
depends on a choice of convention.  Here we have adopted a particular
convention so that all the renormalization effects, including the
renormalization of the squared mass and wave function, are included in
the renormalization of the interaction vertices $\{\V_{2n}\}$.
\item The constant $y_{2n}$ defined by Eq.~(\ref{scaledim}) is the
scale dimension of the vertex $\V_{2n}$.  For example, we find
\begin{equation}
y_2 = 2,\quad y_4 = 0,\quad y_6 = -2,\quad y_8 = -4,\quad \cdots
\end{equation}
The left-hand side of the ERG equation (\ref{diffeq}) implies that the
effect of $\V_{2n} (0)$ on the vertices $\{\V_{2n} (t)\}$ at the
logarithmic scale $t$ is of order $\e^{y_{2n}t}$.  Hence, $\V_2$ is
relevant, $\V_4$ is marginal, and $\V_{2n \ge 6}$ are irrelevant.
This point will be explained again at the end of sect.~\ref{integral}.
\end{enumerate}

\subsection{Conventional use of the ERG equations}

In Ref.~\onlinecite{pol84} the ERG equation (\ref{polchinski}) was
used to prove the perturbative renormalizability of the $\phi^4$
theory.  As the starting point of a renormalization group flow we
choose a bare theory defined by the squared mass $m^2 \e^{2 t_0}$ and
the vertices
\begin{eqnarray}
\V_2 (t_0; p) &=& a_2 (t_0; \lambda) + m^2 \e^{2 t_0} z_m (t_0;
\lambda) + p^2 z_\phi (t_0; \lambda) \\ \V_4 (t_0; p_1,\cdots, p_4)
&=& (- \lambda) \left[ 1 + z_\lambda (t_0; \lambda)
\right]\\ \V_{2n \ge 6} (t_0; p_1,\cdots,p_{2n}) &=& 0
\end{eqnarray}
where $t_0$ is a large \textbf{negative} constant, and $a_2, z_m,
z_\phi$, and $z_\lambda$ are perturbative series in the coupling
constant $\lambda$.  We run the ERG to obtain $\{\V_{2n} (t=0)\}$.  If
$\{\V_{2n} (0)\}$ exist in the limit $t_0 \to - \infty$, we call the
theory renormalizable.  Polchinski showed that the limit exists if we
choose $a_2, z_m, z_\phi$, and $z_\lambda$ as appropriate power series
in $\lambda$ and $t_0$, and that the approach to the limit behaves as
$\e^{2t_0}$ with power corrections in $t_0$ at each order of
perturbation theory.

In the next section we will introduce a more direct way of obtaining
the continuum limit $\{\V_{2n}(0)\}$.  We will not define the
continuum limit by taking the infrared limit of a bare theory as
above, but we will define it in terms of integral equations that the
continuum limit must obey.  The integral equations construct
$S(\infty)$ without the help of any bare theory.

\section{\label{integral}Construction of integral equations}

We construct integral equations from the differential equation
(\ref{polchinski}) (or equivalently Eqs.~(\ref{diffeq})) by following
a standard procedure.  An integral equation is a combination of a
differential equation with an initial condition, and in our case it is
the ultraviolet asymptotic behavior of the vertex functions that plays
the role of the initial condition.

If the theory had a good honest ultraviolet fixed point, the
asymptotic behavior of the vertex functions would be simply
\begin{equation}
\V_{2n} (- t; p_1,\cdots,p_{2n}) \longrightarrow \V_{2n}^*
(p_1,\cdots,p_{2n})\qquad \text{as}\quad t \to + \infty
\end{equation}
where $\{\V_{2n}^*\}$ are the fixed-point vertices.  The perturbative
$\phi^4$ theory does not have an ultraviolet fixed point, and we must
replace the above asymptotic conditions by alternative conditions.  We
impose that the vertices be given by
\begin{equation}
\V_{2n} (- t; p_1,\cdots,p_{2n}) \longrightarrow A_{2n} (-t; p_1,
\cdots, p_{2n}) + m^2 \e^{-2t} B_{2n} (-t; p_1, \cdots, p_{2n}) +
\cdots \label{asymp}
\end{equation}
as $t \to + \infty$ where
\begin{enumerate}
\item The right-hand side is an expansion in powers of $m^2
\e^{-2t}$.
\item $A_{2n}$ and $B_{2n}$, independent of $m^2$, are finite order
polynomials of $t$ at each order in perturbation theory.
\item $A_{2n}$ and $B_{2n}$ are local, i.e., they can be expanded in
powers of momenta if the momenta are small compared to $1$.
\end{enumerate}
We will construct integral equations using the above assumptions, and
in sect.~\ref{solution} we will justify the assumptions using the
integral equations themselves.

Here is a comment on the sign convention for the parameter $t$.  The
parameter $t$ was introduced in the previous section to denote the
logarithmic renormalization scale.  It grows as we go downstream
toward infrared on the RG flow.  Since we will need to go upstream to
write down integral equations, we will mainly deal with negative $t$
in the rest of the paper.  Since we easily forget that $t$ is
negative, we denote it explicitly as $-t$ so that $t > 0$ when we go
upstream on the RG flow.

The asymptotic behavior (\ref{asymp}) implies in particular
\begin{eqnarray}
&&\e^{2t} \V_2 (- t; p \e^{-t}) \to \e^{2t} A_2 (-t; 0) + p^2
\frac{d}{dp^2} A_2 (-t; p)\Bigg|_{p^2=0} + m^2 B_2 (-t; 0)
\label{V2asymp}\\ &&\V_4 (-t; p_1 \e^{-t},\cdots,p_4 \e^{-t}) \to A_4
(-t; 0,0,0,0) \label{V4asymp}\\ &&\e^{y_{2n} t} \V_{2n} (-t; p_1
\e^{-t},\cdots,p_{2n}\e^{-t}) \to 0 \quad \text{for}\quad 2n \ge 6
\end{eqnarray}
as $t \to + \infty$ where the corrections are suppressed by $\e^{-2t}$
(with powers of $t$).  Here we recall that the squared mass that goes
with the vertices $\{\V_{2n} (-t)\}$ is $m^2 \e^{-2t}$, and the above
expansions are taken in powers of $m^2 \e^{-2t}$ and momenta $p
\e^{-t}$.  The last equation is valid because of $y_{2n} < 0$ and the
assumed polynomial behavior of $\V_{2n}(-t)$.

The above asymptotic behavior makes the following equations trivially
valid:
\begin{eqnarray}
&&\e^{2t} \V_2 (-t; p \e^{-t}) \nonumber\\
&=& \lim_{T \to \infty} \Bigg[
\left(\e^{2t} \V_2 (-t; p \e^{-t}) - \e^{2(t+T)} \V_2 (-t-T; p
\e^{-t-T})\right) \nonumber\\
&&\qquad\qquad + \e^{2(t+T)} A_2 (-t-T; 0) + p^2 \frac{d}{dp^2} A_2
(-t-T; p)\Bigg|_{p^2=0} + m^2 B_2 (-t-T;0) \,\Bigg]\\ &&\V_4 (-t; p_1
\e^{-t},\cdots, p_4 \e^{-t})\nonumber\\
&=& \lim_{T \to \infty} \Bigg[ \left( \V_4 (-t;
p_1 \e^{-t},\cdots,p_4 \e^{-t}) - \V_4 (-t-T; p_1 \e^{-t-T},\cdots,
p_4 \e^{-t-T}) \right) \nonumber\\
&&\qquad\qquad\qquad+ A_4 (-t-T; 0,0,0,0) \Bigg]\\
&& \e^{y_{2n} t} \V_{2n} (-t; p_1 \e^{-t},\cdots,p_{2n} \e^{-t})\\
&=&\lim_{T \to \infty} \left[ \e^{y_{2n} t} \V_{2n} (-t; p_1
\e^{-t},\cdots, p_{2n} \e^{-t}) - \e^{y_{2n} (t+T)} \V_{2n} (-t-T; p_1
\e^{-t-T},\cdots, p_{2n} \e^{-t-T}) \right]\nonumber
\end{eqnarray}
The differences of the vertices on the right-hand sides are obtained
by integrating the differential ERG equations (\ref{diffeq}) of the
previous section.  We obtain
\begin{eqnarray}
&&\e^{2t} \V_2 (-t; p \e^{-t})\nonumber\\ &=& \lim_{T \to \infty}
\Bigg[\, \int_0^T dt'\, \Bigg\{ \e^{2(t+t')} \V_2 (-t-t'; p
\e^{-t-t'}) \frac{\Delta (p \e^{-t-t'})}{p^2 + m^2} \e^{2(t+t')} \V_2
(-t-t'; p \e^{-t-t'}) \nonumber\\ &&\qquad + \frac{1}{2} \int_q
\frac{\Delta (q \e^{-t-t'})}{q^2+m^2}\, \V_4 (-t-t'; q \e^{-t-t'}, -q
\e^{-t-t'}, p \e^{-t-t'}, - p \e^{-t-t'}) \Bigg\} \nonumber\\ &&\qquad
+ \e^{2(t+T)} A_2 (-t-T) + p^2 C_2 (-t-T) + m^2 B_2 (-t-T) \,\Bigg]\\
&&\V_4 (-t; p_1 \e^{-t},\cdots, p_4 \e^{-t})\nonumber\\ &=& \lim_{T
\to \infty} \Bigg[\, \int_0^T dt'\, \Bigg\{ \sum_{i=1}^4 \e^{2(t+t')}
\V_2 (-t-t'; p_i \e^{-t-t'}) \, \frac{\Delta (p_i \e^{-t-t'})}{p_i^2 +
m^2} \nonumber\\ &&\qquad\qquad\qquad\qquad\qquad\qquad \times \, \V_4
(-t-t'; p_1 \e^{-t-t'},\cdots, p_4 \e^{-t-t'})\nonumber\\ &&\qquad +
\frac{1}{2} \int_q \frac{\Delta (q \e^{-t-t'})}{q^2 + m^2}\, \e^{- 2
(t+t')} \V_6 (-t-t'; q \e^{-t-t'}, -q \e^{-t-t'}, p_1 \e^{-t-t'},
\cdots, p_4 \e^{-t-t'})\, \Bigg\}\nonumber\\ &&\qquad + A_4 (-t-T)
\,\Bigg]\\ &&\e^{y_{2n} t} \V_{2n} (-t; p_1 \e^{-t},\cdots, p_{2n}
\e^{-t})\nonumber\\ &=& \lim_{T \to \infty} \int_0^T dt'\, \Bigg\{
\sum_{k=0}^{\left[\frac{n-1}{2}\right]} \sum_{\mathrm{partitions:}
\atop I + J = \{2n\}} \e^{y_{2(k+1)}(t+t')} \V_{2(k+1)} (-t-t'; p_I
\e^{-t-t'})\nonumber\\ &&\qquad\qquad\qquad\qquad \times \frac{\Delta
(p_I \e^{-t-t'})}{p_I^2 + m^2}\,\e^{y_{2(n-k)}(t+t')} \V_{2(n-k)}
(-t-t'; p_J \e^{-t-t'}) \nonumber\\ &&\, + \frac{1}{2} \int_q
\frac{\Delta (q \e^{-t-t'})}{q^2+m^2}\, \e^{y_{2(n+1)} (t+t')}
\V_{2(n+1)} (-t-t'; q \e^{-t-t'}, - q \e^{-t-t'}, p_1
\e^{-t-t'},\cdots )\, \Bigg\}
\end{eqnarray}
where we have introduced the notation
\begin{eqnarray}
A_2 (-t) &\equiv& A_2 (-t; 0),\quad B_2 (-t) \equiv B_2 (-t;0),\quad
C_2 (-t) \equiv \frac{\partial}{\partial p^2} A_2 (-t;
p)\Bigg|_{p^2=0}\\ A_4 (-t) &\equiv& A_4 (-t; 0,0,0,0)
\end{eqnarray}
and the short-hand notations $p_I, p_J$ have been used.

The above integral equations are not self-contained yet, since the
right-hand sides depend on the asymptotic forms $A_2, B_2, C_2$, and
$A_4$ which are known only after the left-hand sides, i.e., $\V_2$ and
$\V_4$, are known.

A crucial observation is to be made now: the requirement that the
limits $T \to + \infty$ to exist for the above equations determines
the asymptotic forms.  We will explain this in the remainder of this
section.

For large $T$, we find the following asymptotic behavior using
(\ref{asymp}):
\begin{eqnarray}
&&\frac{1}{2} \int_q \frac{\Delta (q \e^{-t-T})}{q^2 + m^2} \V_4
(-t-T; q \e^{-t-T}, - q \e^{-t-T}, p \e^{-t-T}, -p
\e^{-t-T})\nonumber\\ &&\longrightarrow \frac{1}{2} \,\e^{2(t+T)}
\int_q \frac{\Delta (q)}{q^2} A_4 (-t-T; q, -q, 0,0)\nonumber\\
&&\qquad\quad + p^2 \frac{1}{2} \left( \frac{d}{dp^2} \int_q
\frac{\Delta (q)}{q^2} \, A_4 (-t-T;q,-q,p,-p) \right)_{p^2=0}
\nonumber\\ &&\qquad\quad + m^2 \frac{1}{2} \int_q \Delta (q) \left(
\frac{1}{q^2} B_4 (-t-T; q,-q,0,0) - \frac{1}{q^4} A_4 (-t-T; q, -q,
0,0) \right), \\ &&\frac{1}{2} \int_q \frac{\Delta (q \e^{-t-T})}{q^2 +
m^2} \e^{-2(t+T)} \V_6 (-t-T; q \e^{-t-T}, -q \e^{-t-T}, p_1
\e^{-t-T}, \cdots, p_4 \e^{-t-T} )\nonumber\\ &&\longrightarrow
\frac{1}{2} \int_q \frac{\Delta (q)}{q^2} A_6 (-t-T; q, -q, 0, 0,0,0)
\end{eqnarray}
In deriving this it is important to note that $\Delta (q)$ is
nonvanishing only for $1 < |q| < 2$.  

The above asymptotic behavior determines the $t$-dependence of the
asymptotic forms $A_2, B_2, C_2$, and $A_4$ so that the limit $T \to
+\infty$ of the integral equations exist.  We must therefore obtain
\begin{eqnarray}
- \frac{d}{dt} \left( \e^{2t} A_2 (-t) \right) &=& \e^{2t}\,
\frac{1}{2} \int_q \, \frac{\Delta (q)}{q^2}\, A_4 (-t; q, -q, 0,0)
\label{dA2dt}\\ - \frac{d}{dt} B_2 (-t) &=& \frac{1}{2} \int_q
\,\Delta (q) \left( \frac{1}{q^2} B_4 (-t; q, -q, 0,0) - \frac{1}{q^4}
A_4 (-t; q, -q, 0, 0) \right) \label{dB2dt}\\ - \frac{d}{dt} C_2 (-t)
&=& \frac{1}{2} \frac{\partial}{\partial p^2} \int_q \, \frac{\Delta
(q)}{q^2} A_4 (-t; q,-q, p, -p)\Bigg|_{p^2=0}\label{dC2dt}\\ -
\frac{d}{dt} A_4 (-t) &=& \frac{1}{2} \int_q \, \frac{\Delta
(q)}{q^2}\, A_6 (-t; q, -q, 0,0,0,0) \label{dA4dt}
\end{eqnarray}
Thus, the asymptotic forms are determined by the asymptotic forms of
the higher point vertices up to $t$-independent constants.  Hence, we
obtain
\begin{eqnarray}
\e^{2t} A_2 (-t) &=& - \int^t dt'\, \e^{2 t'} \, \frac{1}{2} \int_q \,
\frac{\Delta (q)}{q^2}\, A_4 (-t'; q, -q, 0,0)\label{A2}\\ B_2 (-t)
&=& \int_0^t dt'\, \frac{1}{2} \int_q \,\Delta (q) \left( -
\frac{1}{q^2} B_4 (-t'; q, -q, 0,0) + \frac{1}{q^4} A_4 (-t'; q, -q,
0, 0) \right)\nonumber\\ &&\quad + B_2 (0) \label{B2}\\ C_2 (-t) &=& -
\int_0^t dt'\, \frac{1}{2} \frac{d}{dp^2} \int_q \, \frac{\Delta
(q)}{q^2} A_4 (-t'; q,-q, p, -p)\Bigg|_{p^2=0} + C_2 (0) \label{C2}\\
A_4 (-t) &=& - \int_0^t dt'\, \frac{1}{2} \int_q \, \frac{\Delta
(q)}{q^2}\, A_6 (-t; q, -q, 0,0,0,0) + A_4 (0) \label{A4}
\end{eqnarray}
Here, $B_2 (0), C_2(0)$, and $A_4 (0)$ are $t$-independent constants
which cannot be determined by the differential equations
(\ref{dB2dt},\ref{dC2dt},\ref{dA4dt}).  The constants $B_2(0), C_2(0)$
have to do with finite renormalization of the squared mass and wave
function, respectively.  The constant $A_4(0)$ is the self-coupling
constant, and together with $m^2$ it parametrizes the space of
continuum limit $S(\infty)$.  We will discuss more about these finite
constants in sect.~\ref{univ}.

The determination of $A_2 (-t)$ by Eq.~(\ref{A2}) needs an
explanation.  As it is, the integral over $t'$ is ambiguous by a
constant, which implies that $A_2 (-t)$ is ambiguous by a constant
multiple of $\e^{-2t}$.  From Eq.~(\ref{V2asymp}) the large $t$
behavior of the two-point vertex at zero momentum is given by
\begin{equation}
\V_2 (-t;0) \longrightarrow A_2 (-t) + m^2 \e^{-2t} B_2 (-t)
\end{equation}
Hence, the ambiguity of order $\e^{-2t}$ in $A_2 (-t)$ has the same
order of magnitude as the $B_2 (-t)$ term.  We wish to remove the
ambiguity in such a way that only the term proportional to $m^2$ gives
the order $\e^{-2t}$ contribution to the asymptotic form of $\V_2
(-t;0)$ above.  This choice is equivalent to the mass independent
scheme, and it turns out that with this choice the massless theory is
given by $m^2 = 0$.\footnote{For $m^2=0$ to correspond to a physically
massless theory, it is important to take the momentum cutoff function
$K(p)$ independent of $m^2$.  It is the well-known fine tuning problem
of the $\phi^4$ theory that the $m^2$ term is exponentially small
compared to the $A_2 (-t)$ term.}  To complete the definition of $A_2
(-t)$, we must first define a $k$-th order polynomial $T_k (t)$ by the
condition
\begin{equation}
\frac{d}{dt} \left( \e^{2t} T_k (t) \right) = \e^{2t} t^k
\end{equation}
Imposing that $T_k (t)$ be a polynomial, we have removed the potential
ambiguity of order $\e^{-2t}$.  Now, given a power series expansion in
$t$
\begin{equation}
\frac{1}{2} \int_q \frac{\Delta (q)}{q^2}\, A_4 (-t; q,-q,0,0) =
\sum_{k=0}^\infty t^k P_k
\end{equation}
we define $A_2 (-t)$ unambiguously by
\begin{equation}
A_2 (-t) \equiv - \sum_{k=0}^\infty T_k (t) \,P_k \label{A2precise}
\end{equation}
This is the precise meaning of Eq.~(\ref{A2}).  For a concrete
expression of the polynomial $T_k (t)$, please refer to Appendix
\ref{Tk}.

We have thus obtained the following integral equations:
\begin{eqnarray}
&&\e^{2t} \V_2 (-t; p \e^{-t})\nonumber\\ &=& \int_0^\infty dt'\,
\Bigg[\, \e^{2(t+t')} \V_2 (-t-t'; p \e^{-t-t'}) \frac{\Delta (p
\e^{-t-t'})}{p^2 + m^2} \e^{2(t+t')} \V_2 (-t-t'; p \e^{-t-t'})
\nonumber\\ &&\, + \frac{1}{2} \int_q \,\Delta (q \e^{-t-t'}) \Bigg\{
\, \frac{1}{q^2 + m^2} \V_4 (-t-t'; q \e^{-t-t'}, -q \e^{-t-t'}, p
\e^{-t-t'}, - p \e^{-t-t'}) \nonumber\\ &&\qquad - \frac{1}{q^2} A_4
(-t-t'; q \e^{-t-t'}, - q \e^{-t-t'}, 0, 0) \nonumber\\ && \qquad -
p^2 \e^{-2(t+t')} \frac{1}{q^2} \frac{\partial}{\partial p^2} A_4
(-t-t'; q \e^{-t-t'}, - q \e^{-t-t'}, p, -p)\Bigg|_{p^2=0}\nonumber\\
&&\qquad - m^2 \e^{-2(t+t')} \Bigg( \frac{1}{q^2} B_4 (-t-t'; q
\e^{-t-t'}, - q \e^{-t-t'},0,0) \nonumber\\ &&\qquad\qquad\qquad\qquad
- \frac{1}{q^4} A_4 (-t-t'; q \e^{-t-t'}, - q \e^{-t-t'}, 0, 0) \Bigg)
\; \Bigg\} \; \Bigg]\nonumber\\ && + A_2 (-t) + p^2 \e^{-2t} C_2 (-t) +
m^2 \e^{-2t} B_2 (-t)\, , \label{inteq2}\\ &&\V_4 (-t; p_1
\e^{-t},\cdots, p_4 e^{-t})\nonumber\\ &=& \int_0^\infty dt'\, \Bigg[
\, \sum_{i=1}^4 \e^{2(t+t')} \V_2 (-t-t'; p_i \e^{-t-t'})\,\frac{\Delta
(p_i \e^{-t-t'})}{p_i^2 + m^2} \,\V_4 (-t-t'; p_1 \e^{-t-t'},\cdots,
p_4 \e^{-t-t'})\nonumber\\ &&\quad + \frac{1}{2} \int_q \Delta (q
\e^{-t-t'}) \Bigg\{ \frac{1}{q^2+m^2} \e^{- 2 (t+t')} \V_6 (-t-t'; q
\e^{-t-t'}, -q \e^{-t-t'}, p_1 \e^{-t-t'}, \cdots, p_4 \e^{-t-t'})\,
\nonumber\\ &&\qquad\qquad - \frac{1}{q^2} \e^{-2(t+t')} A_6 (-t-t'; q
\e^{-t-t'}, -q \e^{-t-t'}, 0,\cdots,0) \, \Bigg\} \; \Bigg] \nonumber\\
&&+ A_4 (-t)\, , \label{inteq4}\\ &&\e^{y_{2n} t} \V_{2n} (-t; p_1
\e^{-t},\cdots, p_{2n} \e^{-t})\nonumber\\ &=& \int_0^\infty dt'\,
\Bigg\{ \sum_{k=0}^{\left[\frac{n-1}{2}\right]}
\sum_{\mathrm{partitions:} \atop I + J = \{2n\}} \e^{y_{2(k+1)}(t+t')}
\V_{2(k+1)} (-t-t'; p_I \e^{-t-t'})\nonumber\\
&&\qquad\qquad\qquad\qquad \times \frac{\Delta (p_I \e^{-t-t'})}{p_I^2
+ m^2}\,\e^{y_{2(n-k)}(t+t')} \V_{2(n-k)} (-t-t'; p_J \e^{-t-t'})
\nonumber\\ &&\, + \frac{1}{2} \int_q \frac{\Delta (q
\e^{-t-t'})}{q^2+m^2}\, \e^{y_{2(n+1)} (t+t')} \V_{2(n+1)} (-t-t'; q
\e^{-t-t'}, - q \e^{-t-t'}, p_1 \e^{-t-t'},\cdots )\, \Bigg\}
\label{inteq2n}
\end{eqnarray}
where $A_2 (-t)$ is given by Eqs.~(\ref{A2},\ref{A2precise}), $B_2
(-t)$ by (\ref{B2}), $C_2 (-t)$ by (\ref{C2}), and $A_4 (-t)$ by
(\ref{A4}).  The constants $B_2 (0), C_2 (0)$, and $A_4 (0)$ are input
parameters to be discussed further in sect.~\ref{univ}.  These
integral equations are self-contained in the sense that they admit a
perturbative solution as explained in the next section.

Before we end this section, let us make two observations.  We first
observe how the relevance, marginality, and irrelevance of vertices
manifest themselves in the above integral equations.  We notice that
the $2n$-point vertex $\V_{2n} (-t-t')$ always appears multiplied by
the exponential factor $\e^{y_{2n}(t+t')}$.  Thus, at scale $-t$ the
effect of the two-point vertex $\V_2 (-t-t')$ at scale $-t-t'$ is of
order $\e^{2t'} \gg 1$ if $t' \gg 1$, and it is relevant.  The effect
of the four-point vertex $\V_4 (-t-t')$ at scale $-t$ is unsuppressed
or marginal.  But the effect of $\V_{2n \ge 6} (-t-t')$ is only of
order $\e^{y_{2n}t'} \ll 1$, and it is irrelevant.

We also observe the mechanism behind the finiteness of the integrals
over $t'$ in the integral equations.  The right-hand sides of the
integral equations have two parts.  The first part consists of
products of two vertices.  For any external momentum $p$, $\Delta (p
\e^{-t'})$ vanishes for large $t'$ since $\Delta (p) = 0$ for $|p| <
1$.  Hence, the first part is finite upon integration over $t'$.  The
second part consists of a loop integral over the momentum $q$.  For
large $t'$ the integrand of the $t'$ integral behaves as $\e^{- 2 t'}$
either by the exponential factor $\e^{y_{2(n+1)}t'}$ or by the
subtractions of asymptotic forms.  Thus, the second part is also
finite upon integration over $t'$.

\section{\label{solution}Solution of integral equations}

In this section we wish to show that the integral equations
(\ref{inteq2}--\ref{inteq2n}) derived in the previous section
determine the vertex functions order by order in perturbation theory.

\subsection{Flow of perturbative solutions}

To solve the integral equations (\ref{inteq2}--\ref{inteq2n}) for the
vertex functions $\{\V_{2n} (t=0)\}$ at scale $t=0$, we must determine
the vertex functions along the entire renormalization group flow from
$t=+\infty$ leading up to the end point $t=0$.  Each renormalization
flow is parametrized by the squared mass $m^2$, and the constants $B_2
(0), C_2 (0)$, and $A_4 (0)$.  In order to solve the integral
equations perturbatively in powers of the coupling constant $\lambda$,
we must assume that these constants can be expanded in powers of
$\lambda$ as
\begin{eqnarray}
B_2 (0) &=& \sum_{k=1}^\infty (-\lambda)^k z_m^{(k)}\label{B2zero}\\
C_2 (0) &=& \sum_{k=1}^\infty (-\lambda)^k z_\phi^{(k)}\label{C2zero}\\
A_4 (0) &=& - \lambda + \sum_{k=1}^\infty (-\lambda)^{1+k}
z_\lambda^{(k)} \label{A4zero}
\end{eqnarray}
where $z_m^{(k)}, z_\phi^{(k)}$, and $z_\lambda^{(k)}$ are arbitrary
constants.  The choice of these constants correspond to a convention
or a renormalization scheme as will be discussed more fully in
sect.~\ref{univ}.  One choice convenient for explicit calculations is
our version of the ``minimal subtraction'' scheme defined by
\begin{equation}
B_2 (0) = C_2 (0) = 0,\quad A_4 (0) = - \lambda \label{MS}
\end{equation}
In the following discussion, however, we will not choose the minimal
subtraction scheme, and we will keep our choice of $B_2 (0), C_2 (0),
A_4 (0)$ arbitrary.\footnote{The problem of the minimal subtraction
scheme (\ref{MS}) is that it is not RG invariant.  $B_2 (0) = C_2 (0)
= 0$ does not imply $B_2 (-t) = C_2 (-t) = 0$ for arbitrary $t$.  To
introduce an RG invariant MS scheme, we must modify our ERG equation.}

Let us recall the recursive solution of an integral equation
\begin{equation}
f(x) = \lambda + \int dy\, G(x, y) f(y)^2
\end{equation}
where $G$ is a known integration kernel.  The recursive solution gives
$f(x)$ as a power series in $\lambda$:
\begin{equation}
f(x) = \lambda + \sum_{k=1}^\infty \lambda^{1+k} f_k (x)
\end{equation}
If $f(x)$ is determined to order $\lambda^{n-1}$, we can use it to
compute the right-hand side up to order $\lambda^n$.  Thus, $f(x)$ is
obtained to order $\lambda^n$.  The recursive method works because the
integral is quadratic in $f$.

The structure of our integral equations (\ref{inteq2}--\ref{inteq2n})
is similar to the simple integral equation above.  The starting point
of the perturbative calculations of the vertices is the four-point
vertex $\V_4$ at order $\lambda$:
\begin{equation}
\V_4 (-t; p_1, \cdots, p_4) = A_4 (-t) = - \lambda
\end{equation}
Everything bootstraps from this.

Let us briefly sketch the perturbative procedure leaving details to
the next subsection.  (Lowest order calculations are given in Appendix
\ref{calculations}.)  Suppose we have computed all the vertices up to
order $\lambda^{n-1}$ at which only $\V_2, \cdots, \V_{2n}$ are
non-vanishing.  At order $\lambda^n$ ($n \ge 1$), we must start from
$\V_{2(n+1)}$ which is given explicitly by the tree-level Feynman
diagrams with $n$ vertices $(- \lambda)$ and $n-1$ internal
propagators
\[
\frac{1 - K(p)}{p^2 + m^2 \e^{-2t}}
\]
For example, we find
\begin{eqnarray}
&&\e^{-2t} \V_6 (-t; p_1 \e^{-t} ,\cdots,p_6 \e^{-t})\nonumber\\ &&
\quad = (-\lambda)^2 \left( \frac{1 -
K((p_1+p_2+p_3)\e^{-t})}{(p_1+p_2+p_3)^2 + m^2} + \text{9
permutations} \right)\\ &&\e^{-4t} \V_8 (-t; p_1 \e^{-t},\cdots, p_8
\e^{-t})\nonumber\\ && \quad = (-\lambda)^3 \Bigg( \frac{1 - K((p_1 +
p_2 + p_3)\e^{-t})}{(p_1+p_2+p_3)^2 + m^2} \cdot \frac{1 - K((p_4 +
p_5 + p_6)\e^{-t})}{(p_4+p_5+p_6)^2 + m^2} \nonumber\\
&&\qquad\qquad\qquad + \,\text{279 permutations} \Bigg)
\end{eqnarray}
We can obtain these by solving the tree-level integral equations
\begin{eqnarray}
&&\e^{y_{2n}t} \V_{2n} (-t; p_1 \e^{-t},\cdots, p_{2n}
\e^{-t})\nonumber\\ &=& \int_0^\infty dt'\, \Bigg\{
\sum_{k=0}^{\left[\frac{n-1}{2}\right]} \sum_{\mathrm{partitions:}
\atop I + J = \{2n\}} \e^{y_{2(k+1)}(t+t')} \V_{2(k+1)} (-t-t'; p_I
\e^{-t-t'})\nonumber\\ &&\qquad\qquad\qquad\qquad \times \frac{\Delta
(p_I \e^{-t-t'})}{p_I^2 + m^2}\,\e^{y_{2(n-k)}(t+t')} \V_{2(n-k)}
(-t-t'; p_J \e^{-t-t'}) \, \Bigg\}
\end{eqnarray}
where there is no loop integral over $q$ on the right-hand side.  The
dependency of the vertices is such that we need only the tree-level
vertices $\V_4, \cdots, \V_{2n}$ to construct $\V_{2(n+1)}$ at
tree-level.

The next vertex to compute at order $\lambda^n$ is $\V_{2n}$.  The
right-hand side of the integral equation (\ref{inteq2n}) has two
parts.  To get an order $\lambda^n$ contribution from the first part,
we need $\V_{2}$ to order $\lambda$, $\V_4$ to $\lambda^2$, $\V_6$ to
$\lambda^3$, ..., $\V_{2(n-1)}$ to $\lambda^{n-1}$, and $\V_{2n}$ to
order $\lambda^{n-1}$.  All these suffice to be lower order in
$\lambda$ than $\lambda^n$ since the first part consists of a product
of two vertices.  To get an order $\lambda^n$ contribution from the
second part, we need $\V_{2(n+1)}$ to oder $\lambda^n$, which is
obtained by the previous step.  Proceeding analogously we can
calculate $\V_{2(n-1)}, \cdots, \V_2$ up to order $\lambda^n$.

The flow of perturbative calculations sketched above is shown in
Fig.~1.  We will elaborate on this further in the next subsection.
\begin{figure}
\includegraphics{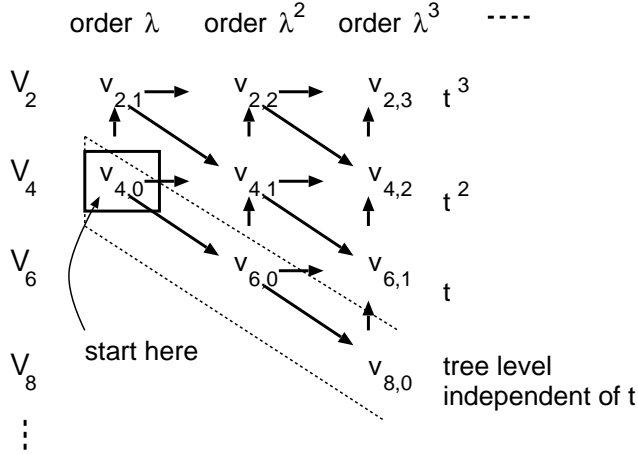}
\caption{Flow of perturbative calculations: every vertex upstream is
necessary to determine a vertex.}
\end{figure}

\subsection{Perturbative proof of the $\lambda, t$-dependence of the vertices}

The purpose of this subsection is to prove the existence of a
perturbative solution to the integral equations
(\ref{inteq2}--\ref{inteq2n}) by proving the following
$\lambda, t$-dependence of the vertex functions:
\begin{equation}
\V_{2n} (-t; p_1, \cdots, p_{2n}) = \sum_{k=0}^\infty
(-\lambda)^{n-1+k} \,v_{2n,k} (-t; p_1,\cdots, p_{2n}; m^2 \e^{-2t})
\label{hypothesis}
\end{equation}
where $v_{2n,k}(-t; p_1, \cdots, p_{2n}; m^2 \e^{-2t})$ is an order
$k$ polynomial of $t$.  $v_{2n,k}$ corresponds to the $k$-loop
contribution to the vertex.  The only exception to
Eq.~(\ref{hypothesis}) is for $n=1$, for which we take the $\lambda$
independent part vanishing:
\begin{equation}
v_{2,0} = 0
\end{equation}
Hence, for $\lambda = 0$, all the vertices $\V_{2n}$ vanish.  The
starting point of the perturbative solution is given by
\begin{equation}
v_{4,0} = 1 \label{start}
\end{equation}
which is independent of the mass and momenta.

We note that by proving the above $\lambda, t$-dependence we also
prove the assumption on the polynomial behavior of the asymptotic
forms (\ref{asymp}).  Eq.~(\ref{hypothesis}) gives
\begin{eqnarray}
A_{2n} (-t; p_1,\cdots,p_{2n}) &=& \sum_{k=0}^\infty
(-\lambda)^{n-1+k}\, v_{2n,k} (-t; p_1, \cdots, p_{2n}; 0)\\ B_{2n}
(-t; p_1,\cdots,p_{2n}) &=& \sum_{k=0}^\infty (-\lambda)^{n-1+k} \,
\frac{\partial}{\partial m^2} v_{2n,k} (-t; p_1, \cdots, p_{2n}; m^2)
\Bigg|_{m^2=0}
\end{eqnarray}
From Eqs.~(\ref{V2asymp},\ref{V4asymp}) we also obtain
\begin{eqnarray}
A_2 (-t) &=& \sum_{k=1}^\infty (-\lambda)^k\, v_{2,k} (-t; 0,0;0)\\
B_2 (-t) &=& \sum_{k=1}^\infty (-\lambda)^k\, \frac{\partial}{\partial
m^2} v_{2,k} (-t; 0,0;m^2) \Bigg|_{m^2=0}\\ C_2 (-t) &=&
\sum_{k=1}^\infty (-\lambda)^k\, \frac{\partial}{\partial p^2} 
v_{2,k} (-t; p,-p;0) \Bigg|_{p^2=0}\\ A_4 (-t) &=& - \lambda +
\sum_{k=1}^\infty (-\lambda)^{1+k} \,v_{4,k} (-t; 0,0,0,0;0)
\end{eqnarray}

The inductive proof of the $\lambda, t$-dependence (\ref{hypothesis})
is straightforward.  The dependence is valid for the starting point
(\ref{start}) of induction.  We wish to prove the validity of the
$\lambda, t$-dependence (\ref{hypothesis}) for $v_{2n,k}$ assuming its
validity for all $v_{2n', k'}$ upstream in Fig.~1 where either
\begin{equation}
n' + k' < n + k 
\end{equation}
or 
\begin{equation}
n' + k' = n + k\quad \text{and}\quad k' < k
\end{equation}
(In Fig.~1, each column has the same $n+k$.  As we go toward right, $n+k$
increases.  As we go up, $k$ increases, and $n$ decreases.)  There are
three cases we must consider separately: $n > 2$, $n = 2$, and $n =
1$.  First we consider the case $n > 2$.  By substituting the assumed
results into the right-hand side of the integral equation
(\ref{inteq2n}) for $\V_{2n}$, we obtain
\begin{eqnarray}
&&v_{2n,k} (-t; p_1, \cdots, p_{2n}; m^2)\nonumber\\
&=& \int_0^\infty dt'\, \Bigg[ \,\e^{(y_{2n}+2)t'}
\sum_{j=0}^{\left[\frac{n-1}{2}\right]} \sum_{l=0}^k
\sum_{\mathrm{partitions:}\atop I+J=\{2n\}} v_{2(j+1),l} (-(t+t'); p_I
\e^{-t'}; m^2 \e^{-2 t'}) \nonumber\\
&&\qquad\qquad\qquad\qquad \times \frac{\Delta(p_I \e^{-t'})}{p_I^2 + m^2} \,
v_{2(n-j),k-l} (-(t+t'); p_J \e^{-t'}; m^2 \e^{-2t'})\nonumber\\
&&\qquad + \frac{1}{2} \int_q \frac{\Delta (q)}{q^2 + m^2 \e^{-2t'}}
\, \e^{y_{2n} t'} v_{2(n+1), k-1} (-(t+t'); q, -q, p_1
\e^{-t'},\cdots; m^2 \e^{-2t'}) \, \Bigg]
\end{eqnarray}
The right-hand side contains only the lower order vertices for which
the induction hypothesis is assumed valid.  The first sum gives at
most order $t^k$, and the second loop integral gives only $t^{k-1}$.
Hence, $v_{2n,k}$ is an order $k$ polynomial of $t$.

Next we look at the special case $n = 2$.  The integral equation
(\ref{inteq4}) gives
\begin{eqnarray}
&&v_{4,k} (-t; p_1,\cdots,p_4; m^2)\nonumber\\ &=& \int_0^\infty dt'\,
\Bigg[ \, \e^{2t'}\, \sum_{i=1}^4 \sum_{l=1}^{k} v_{2,l} (-(t+t'); p_i
\e^{-t'}, - p_i \e^{-t'}; m^2 \e^{- 2 t'})\, \frac{\Delta (p_i
\e^{-t'})}{p_i^2 + m^2}\nonumber\\ &&\qquad\qquad\qquad\qquad \times
\, v_{4,k-l} (-(t+t'); p_1 \e^{-t'}, \cdots, p_4 \e^{-t'}; m^2
\e^{-2t'})\nonumber\\ &&\quad + \frac{1}{2} \int_q \Bigg\{
\frac{\Delta (q)}{q^2 + m^2 \e^{-2t'}} \, v_{6,k-1} (-(t+t'); q,-q,
p_1 \e^{-t'},\cdots, p_4 \e^{-t'}; m^2 \e^{-2t'})\nonumber\\
&&\qquad\qquad - \frac{\Delta (q)}{q^2}\, v_{6,k-1} (-(t+t');
q,-q,0,0,0,0; 0) \, \Bigg\}\,\Bigg]\nonumber\\ &+& v_{4,k}
(-t;0,0,0,0;0)
\end{eqnarray}
The first term in the integral gives at most order $t^k$, and the
second loop integral at most order $t^{k-1}$.  The last term is
obtained from Eqs.~(\ref{A4},\ref{A4zero}) as
\begin{equation}
v_{4,k} (-t;0,0,0,0;0) = z_\lambda^{(k)} - \int_0^t dt'\, \frac{1}{2}
\int_q \frac{\Delta (q)}{q^2} \, v_{6,k-1} (-t'; q,-q,0,0,0,0;0)
\end{equation}
Since $v_{6,k-1} (-t; q,-q,0,0,0,0;0)$ is a polynomial of order $k-1$
by the induction hypothesis, the above equation implies that
$v_{4,k}(-t;0,0,0,0;0)$ is an order $k$ polynomial.

Finally we consider the case $n=1$.  The integral equation
(\ref{inteq2}) gives
\begin{eqnarray}
&& v_{2,k} (-t; p,-p; m^2)\nonumber\\ &=& \int_0^\infty dt'\, \Bigg[\,
\e^{4 t'} \sum_{l=1}^{k-1} v_{2,l} (-(t+t'); p
\e^{-t'}, -p \e^{-t'}; m^2 \e^{-2t'}) \nonumber\\
&&\qquad\qquad\qquad \times\, \frac{\Delta (p
\e^{-t'})}{p^2 + m^2} \, v_{2,k-l} (-(t+t'); p \e^{-t'}, -p \e^{-t'};
m^2 \e^{-2t'})\nonumber\\ &+& \frac{1}{2} \int_q \Delta(q) \Bigg\{ \,
\frac{1}{q^2 + m^2 \e^{-2t'}} \, \e^{2t'} v_{4,k-1} (-(t+t'); q,-q,p
\e^{-t'},-p \e^{-t'}; m^2 \e^{-2t'})\nonumber\\ &&\qquad\qquad\qquad -
\frac{1}{q^2} \e^{2t'} v_{4,k-1} (-(t+t'); q,-q,0,0;0) \nonumber\\
&&\qquad\qquad\qquad - \frac{1}{q^2} p^2 \frac{\partial}{\partial p^2}
v_{4,k-1} (-(t+t'); q,-q,p,-p;0) \Bigg|_{p^2=0}\nonumber\\
&&\qquad\qquad\qquad - \frac{1}{q^2} m^2 \frac{\partial}{\partial m^2}
v_{4,k-1} (-(t+t'); q,-q,0,0;m^2) \Bigg|_{m^2=0} \nonumber\\
&&\qquad\qquad\qquad + \frac{1}{q^4} m^2 v_{4,k-1} (-(t+t');
q,-q,0,0;0)\,\Bigg\}\,\Bigg]\nonumber\\ &+& v_{2,k} (-t;0,0;0) + p^2
\frac{\partial}{\partial p^2} v_{2,k} (-t; p,-p;0) \Bigg|_{p^2=0} +
m^2 \frac{\partial}{\partial m^2} v_{2,k} (-t; 0,0;m^2) \Bigg|_{m^2=0}
\end{eqnarray}
The first sum gives at most order $t^k$, and the second loop integral
over $q$ gives at most order $t^{k-1}$.  The last line is obtained
from Eqs.~(\ref{A2}--\ref{C2}) and Eqs.~(\ref{B2zero},\ref{C2zero}) as
\begin{eqnarray}
\e^{2t} v_{2,k} (-t;0,0;0) &=& - \int^t dt'\, \e^{2t'}\,\frac{1}{2} \int_q
\frac{\Delta (q)}{q^2} v_{4,k-1} (-t'; q,-q,0,0;0) \label{v2k}\\
\frac{\partial}{\partial p^2} v_{2,k} (-t; p,-p;0) \Bigg|_{p^2=0} &=&
z_\phi^{(k)} - \int_0^t dt' \frac{1}{2} \int_q \frac{\Delta (q)}{q^2}
\frac{\partial}{\partial p^2} v_{4,k-1} (-t';
q,-q,p,-p;0)\Bigg|_{p^2=0}\\ \frac{\partial}{\partial m^2} v_{2,k}
(-t;0,0;m^2)\Bigg|_{m^2=0} &=& z_m^{(k)} - \int_0^t dt' \frac{1}{2}
\int_q \Delta (q) \Bigg( \,\frac{1}{q^2} \frac{\partial}{\partial m^2}
v_{4,k-1} (-t'; q,-q,0,0;m^2)\Bigg|_{m^2=0}\nonumber\\
&&\qquad\qquad\qquad - \frac{1}{q^4} v_{4,k-1} (-t'; q,-q,0,0;0)
\,\Bigg)
\end{eqnarray}
The precise meaning of the integral on the right-hand side of
Eq.~(\ref{v2k}) has been given in the paragraph leading to
Eq.~(\ref{A2precise}): the integral converts $t^j$ into an order $j$
polynomial $T_j (t)$.  The induction hypothesis implies that the
left-hand sides in the above are all at most order $t^k$.  Hence, we
have proven that $v_{2,k}$ is at most order $t^k$.

This concludes the inductive proof of the $\lambda, t$-dependence
given by (\ref{hypothesis}).  We have thus proven the existence of a
perturbative solution to the integral equations
(\ref{inteq2}--\ref{inteq2n}).  Since the integral
equations define a continuum limit directly, we have proven the
perturbative renormalizability of the $\phi^4$ theory at the same
time.  

\section{\label{univ}Universality}

In the previous section we have shown the existence of a perturbative
solution of the ERG integral equations (\ref{inteq2}--\ref{inteq2n}).
In this section we consider two issues related to universality: first
we will count the independent degrees of freedom of the continuum
limit, and second we will consider how the Green functions depend (or
not depend) on the choice of a momentum cutoff function $K(p)$.

We first recall that each solution of the integral equations
(\ref{inteq2}--\ref{inteq2n}) gives an entire trajectory of the ERG
flow in the space $S(\infty)$ of the continuum limit.  Each trajectory
is parametrized by $-t$ which ranges from $-\infty$ to $0$, and it is
specified by a squared mass $m^2$ and three input parameters $B_2
(0)$, $C_2 (0)$, and $A_4 (0)$.  We can regard $m^2$, and $B_2 (0)$,
$C_2 (0)$, $A_4 (0)$ as the four coordinates of the end point of the
ERG trajectory.  Hence, the space $S(\infty)$ is four dimensional.
According to the usual understanding of the $\phi^4$ theory, however,
the continuum limit has only two parameters: a squared mass $m^2$ and
a self-coupling constant $\lambda$.  We wish to reconcile this
discrepancy.

\begin{figure}
\includegraphics{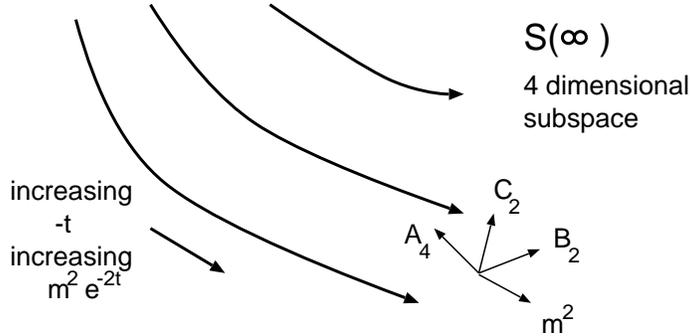}
\caption{Flows of ERG: the perfect actions make a 4-dimensional
subspace with coordinates $m^2$, $B_2(0)$, $C_2(0)$, $A_4(0)$}
\end{figure}

Clearly the parameter $A_4 (0)$ corresponds to the self-coupling
constant $\lambda$.  The other two parameters $B_2 (0)$ and $C_2 (0)$,
which we can take as zero in the minimal subtraction scheme
(\ref{MS}), are related to finite renormalization of the squared mass
and wave function, respectively.  

Since the space of the continuum limit $S(\infty)$ is physically
two-dimensional, there should be a two dimensional group of
transformations which relate physically equivalent theories.  More
concretely, we should be able to find an infinitesimal change of the
parameters $m^2$, $B_2 (0)$, $C_2 (0)$, and $A_4 (0)$ so that the Green
functions remain unchanged up to normalization.  Such a transformation
should map an entire ERG flow to another physically equivalent ERG
flow.  Without derivation, we write down the infinitesimal
transformation $\V_{2n} \to \V_{2n} + \delta \V_{2n}$ with the
expected properties:
\begin{eqnarray}
&&\e^{2t} \delta \V_2 (-t; p \e^{-t}) = \eta (p^2+m^2) +
\ep m^2 \nonumber\\ && \quad + \e^{2t} \V_2 (-t; p \e^{-t})
\left\{ - \eta + 2 (1 - K(p \e^{-t})) \left( \eta + \frac{\ep
m^2}{p^2+m^2} \right) \right\} \nonumber\\ &&\quad - \left( \e^{2t}
\V_2 (-t; p \e^{-t}) \right)^2 \frac{K(p \e^{-t}) (1-K(p
\e^{-t}))}{p^2+m^2} \left( \eta + \frac{\ep m^2}{p^2+m^2} \right)
\nonumber\\ && \quad - \frac{1}{2} \int_q \frac{K(q \e^{-t})(1 - K(q
\e^{-t}))}{q^2 + m^2} \left( \eta + \frac{\ep m^2}{q^2+m^2} \right) \,
\V_4 (-t; q \e^{-t}, - q \e^{-t}, p \e^{-t}, - p \e^{-t}) \label{deltaV2}
\end{eqnarray}
and for $2n \ge 4$
\begin{eqnarray}
&&\e^{y_{2n}t} \delta V_{2n} (-t; p_1 \e^{-t},\cdots, p_{2n} \e^{-t})
\nonumber\\ &=& \sum_{i=1}^{2n} \left\{ - \frac{\eta}{2} + (1 - K(p_i
\e^{-t})) \left( \eta + \frac{\ep m^2}{p_i^2 + m^2} \right)
\right\}\, \e^{y_{2n} t} \V_{2n} (-t; p_1 \e^{-t}, \cdots, p_{2n}
\e^{-t} )\nonumber\\ && \, -
\sum_{k=0}^{\left[\frac{n-1}{2}\right]}
\sum_{\mathrm{partitions:}\atop I + J = \{2n\}} \e^{y_{2(k+1)}t}
\V_{2(k+1)} (-t; p_I \e^{-t})\, \frac{K(p_I \e^{-t}) (1 - K(p_I
\e^{-t}))}{p_I^2 + m^2} \, \left\{ \eta + \frac{\ep m^2}{p_I^2 + m^2}
\right\} \nonumber\\ && \qquad\qquad\qquad\qquad\qquad \times
\,\e^{y_{2(n-k)} t} \V_{2(n-k)} (-t; p_J \e^{-t})\nonumber\\ &&\, -
\frac{1}{2} \int_q \frac{K(q \e^{-t}) (1 - K(q \e^{-t}))}{q^2 + m^2}\,
\left\{ \eta + \frac{\ep m^2}{q^2 + m^2} \right\}\nonumber\\
&&\qquad\qquad\qquad\times\, \e^{y_{2(n+1)} t} \V_{2(n+1)} (-t; q
\e^{-t}, - q \e^{-t}, p_1 \e^{-t}, \cdots, p_{2n} \e^{-t} )\label{deltaV2n}
\end{eqnarray}
where $\ep, \eta$ are infinitesimal constants.  

The above transformation satisfies the following two properties:
\begin{enumerate}
\item The Green functions change only by normalization:
\begin{eqnarray}
&&\vev{\phi (p_1) \cdots \phi (p_{2n-1}) \phi}_{m^2 \e^{-2t}, \V (-t)}
\nonumber\\ &&\quad= (1 + n \eta) \vev{\phi (p_1) \cdots \phi
(p_{2n-1}) \phi}_{m^2 (1 + \ep) \e^{-2t}, (\V+ \delta \V)(-t) }
\label{equivalence}
\end{eqnarray}
\item The transformed vertices $(\V_{2n} + \delta \V_{2n})(-t)$
satisfy the ERG equations (\ref{diffeq}) for the squared mass $m^2 (1
+ \ep) \e^{-2t}$.
\end{enumerate}
(For a proof of the above properties, please refer to
Ref.~\onlinecite{hs03}.)

The infinitesimal transformation defined by
(\ref{deltaV2},\ref{deltaV2n}) corresponds to the following
infinitesimal change of the parameters:
\begin{eqnarray}
&&m^2 \longrightarrow m^2 (1 + \ep)\\ &&B_2 (0) \longrightarrow (1 -
\eta) B_2 (0) + \ep + \eta \nonumber\\ &&\quad- \frac{1}{2} \int_q
K(q)(1-K(q))\, \left( (- \eta + \ep) \frac{1}{q^4} A_4 (0;q,-q,0,0) +
\eta \frac{1}{q^2} B_4 (0; q,-q,0,0) \right)\\ &&C_2 (0)
\longrightarrow (1 - \eta) C_2 (0) + \eta \nonumber\\
&&\qquad\qquad\qquad - \eta \frac{1}{2} \frac{\partial}{\partial p^2}
\int_q \frac{K(q)(1-K(q))}{q^2} A_4 (0;q,-q,p,-p)\Bigg|_{p^2=0}\\
&&A_4 (0) \longrightarrow (1 - 2 \eta) A_4 (0) - \eta \frac{1}{2} \int_q
\frac{K(q)(1-K(q))}{q^2}\, A_6 (0; q,-q,0,0,0,0)
\end{eqnarray}
These infinitesimal transformations generate equivalence classes of
theories, and the space of the equivalence classes is two-dimensional.

In Ref.~\onlinecite{hs03} we will modify the ERG equations by
introducing a running squared mass and an anomalous scale dimension of
the field $\phi$.  With the modification we can no longer take $B_2
(0)$ and $C_2 (0)$ as arbitrary, and the space $S(\infty)$ becomes
two-dimensional.

We now proceed to the next issue.  We recall that universality usually
means that the Green functions of the scalar field $\phi$ is unique up
to normalization of the field.  In other words the Green functions in
the continuum limit do not depend on how the continuum limit is taken.
In the present context universality demands that we get the same Green
functions no matter what momentum cutoff function $K(p)$ we use, as
long as $K(p)$ is $1$ for small $|p|$ and $0$ for large $|p|$.  Under
a change of $K$, the Green functions should change in such a way that
the differences can be compensated by appropriate finite change of the
parameters and normalization of the field.

Let us consider the Green functions computed with the vertices
$\{\V_{2n} (-t)\}$ using a modified propagator $(K+\delta K)(p)/(p^2 +
m^2 \e^{-2t})$, where the infinitesimal change $\delta K(p)$ vanishes
for $|p| < 1$ and for large $|p|$.  The change of the Green functions
due to the modified propagator can be reproduced using the original
cutoff function $K(p)$ but using a different set of vertices
$\{(\V_{2n} + \delta \V_{2n})(-t)\}$:
\begin{equation}
\vev{\phi (p_1) \cdots \phi (p_{2n-1}) \phi}_{K,m^2 \e^{-2t},(\V +
\delta \V) (-t)} = \vev{\phi (p_1) \cdots \phi (p_{2n-1})
\phi}_{K+\delta K, m^2 \e^{-2t}, \V (-t)}
\end{equation}
The change $\{\delta \V_{2n} (-t)\}$ of the vertices necessary for the
above equality is most easily obtained by a diagrammatic
consideration.  By interpreting the $\delta K$ not as part of a
propagator but as part of a vertex, we find that the appropriate
infinitesimal change of the vertices is given by
\begin{eqnarray}
&&\e^{y_{2n} t} \delta \V_{2n} (-t; p_1 \e^{-t}, \cdots, p_{2n}
\e^{-t})\nonumber\\ &=& - \sum_{k=0}^{\left[\frac{n-1}{2}\right]}
\sum_{\mathrm{partitions:}\atop I + J = \{2n\}} \e^{y_{2(k+1)} t}
\V_{2(k+1)} (-t; p_I \e^{-t}) \nonumber\\ &&\qquad\qquad\qquad \times
\frac{\delta K (p_I \e^{-t})}{p_I^2 + m^2}\, \e^{y_{2(n-k)} t}
\V_{2(n-k)} (-t; p_J \e^{-t})\nonumber\\ &&\, - \frac{1}{2} \int_q
\frac{\delta K(q \e^{-t})}{q^2 + m^2}\, \e^{y_{2(n+1)} t} \V_{2(n+1)}
(-t; q \e^{-t}, - q \e^{-t}, p_1 \e^{-t} ,\cdots\, p_{2n} \e^{-t})
\end{eqnarray}
It is straightforward to check that the vertices $\{(\V_{2n} + \delta
\V_{2n})(-t)\}$ satisfy the ERG equations (\ref{diffeq}) with the
squared mass $m^2 \e^{-2t}$.  

The above change of the vertices corresponds to the following change
of the input parameters to the integral ERG equation:
\begin{eqnarray}
\delta B_2 (0) &=& - \frac{1}{2} \int_q \delta K(q) \left( \,
\frac{1}{q^2} B_4 (0; q, -q, 0,0) - \frac{1}{q^4} A_4 (0; q, -q, 0,0) \right)\\
\delta C_2 (0) &=& - \frac{1}{2} \frac{\partial}{\partial p^2} \int_q
\frac{\delta K(q)}{q^2} A_4(0; q,-q,p,-p)\Bigg|_{p^2=0}\\
\delta A_4 (0) &=& - \frac{1}{2} \int_q \frac{\delta K(q)}{q^2} A_6
(0; q, -q, 0,0,0,0)
\end{eqnarray}
Hence, the ERG trajectory specified by $m^2$, $B_2 (0)$, $C_2 (0)$,
and $A_4 (0)$ in $S(\infty)$ with the cutoff $K + \delta K$ is
equivalent to the ERG trajectory specified by $m^2$, $(B_2 + \delta
B_2)(0)$, $(C_2 + \delta C_2)(0)$, and $(A_4 + \delta A_4)(0)$ in
$S(\infty)$ with the cutoff $K$.  Thus, with this equivalence, the
space of theories in the continuum limit is independent of the choice
of a momentum cutoff function $K$.  In other words the continuum limit
is universal.

\section{\label{conclusion}Conclusion}

In this paper we have reformulated the exact renormalization group
equation of Wilson in terms of integral equations.  The advantage of
the integral equations is that they define the continuum limit of a
theory directly.  So far the exact renormalization group has been
studied as differential (or difference) equations, and for
perturbation theory it has been used mainly as a method of
regularization which is particularly convenient for formal studies.
The continuum limit has to be constructed by first introducing a bare
theory and then taking the bare theory to a critical point.  In
comparison the integral equation approach has two advantages: first we
can construct the continuum limit directly, and second the integral
equation naturally provides a self-determining perturbative procedure.

The integral equations are somewhat cumbersome to write down due to
the subtractions necessary for the two- and four-point vertices.
However, the analysis of the structure of the perturbative solution is
straightforward, and the proof of the existence of a perturbative
solution given in sect.~\ref{solution} is one of the simplest proofs
(if not the simplest) of renormalizability of $\phi^4$ theory in the
literature.

Some questions left unanswered in this paper will be answered in a
forthcoming paper.\cite{hs03} In particular it should be interesting
to relate the ordinary renormalization group equations of the
renormalized parameters and fields to the exact renormalization group
equations.  The lowest order results given in Ref.~\onlinecite{hl88}
will be extended to all orders in perturbation theory in
Ref.~\onlinecite{hs03} by modifying the exact renormalization group
equation.

The exact renormalization group has been applied to a wide variety of
theories such as gauge theories, chiral theories, theories with
spontaneous symmetry breaking, supersymmetric theories, and theories
with a real ultraviolet fixed point.  (For example, see
Refs.~\cite{warr88a, warr88b} for applications to gauge, chiral, and
supersymmetric theories.)  We expect that the integral equation
approach introduced in this paper will further simplify the
perturbative studies of those theories.

\appendix
\section{\label{calculations}Lowest order calculations}

We choose the MS scheme:
\begin{equation}
B_2 (0) = C_2 (0) = 0,\quad
A_4 (0) = - \lambda 
\end{equation}

\subsection{Order $\lambda$}

At order $\lambda$ we find
\begin{equation}
\V_4 (-t; p_1 \e^{-t},\cdots, p_4 \e^{-t}) = (-\lambda) \,v_{4,0},\quad
\V_2 (-t; p \e^{-t}) = (-\lambda)\, v_{2,1} (-t)
\end{equation}
where
\begin{eqnarray}
v_{4,0} &=& 1\\ \e^{2t} v_{2,1} (-t) &=& \int_0^\infty dt'\,
\frac{1}{2} \int_q \left[ \, \frac{\Delta (q \e^{-(t+t')})}{q^2 + m^2}
- \frac{\Delta (q\e^{-(t+t')})}{q^2} + m^2 \frac{\Delta (q
\e^{-(t+t')})}{q^4} \,\right]\nonumber\\ &&\quad - \frac{1}{2}
\,\e^{2t} T_0 \int_q \frac{\Delta (q)}{q^2} + t\, m^2\, \frac{1}{2}
\int_q \frac{\Delta (q)}{q^4} \nonumber\\ &=& \frac{1}{2} \int_q
\left[ \frac{1-K(q \e^{-t})}{q^2 + m^2} - \frac{1}{q^2} + (1-K(q
\e^{-t})) \frac{m^2}{q^4} \right] + t\, m^2\, \frac{1}{2} \int_q
\frac{\Delta (q)}{q^4}
\end{eqnarray}
where we used
\begin{equation}
\frac{d}{dt} K(p \e^{-t}) = \Delta (p \e^{-t})
\end{equation}
and $T_0 = 1/2$ is defined by
\begin{equation}
\frac{d}{dt} \left( \e^{2t} T_0 \right) = \e^{2t}
\end{equation}

\subsection{Order $\lambda^2$}

Up to order $\lambda^2$ we find
\begin{eqnarray}
\V_6 (-t; p_1 \e^{-t},\cdots, p_6 \e^{-t}) &=&
(-\lambda)^2 \, v_{6,0} (p_1 \e^{-t},\cdots, p_6 \e^{-t}; m^2
\e^{-2t})\\
\V_4 (-t; p_1 \e^{-t},\cdots, p_4 \e^{-t}) &=& (-\lambda)\, v_{4,0} +
(-\lambda)^2 \, v_{4,1} (-t; p_1 \e^{-t},\cdots, p_4 \e^{-t}; m^2
\e^{-2t})\\
\V_2 (-t; p \e^{-t}) &=& (-\lambda) \,v_{2,1} (-t) + (-\lambda)^2 \,
v_{2,2} (-t; p \e^{-t}, - p \e^{-t}; m^2 \e^{-2t})
\end{eqnarray}
We must start from the six-point function.
\begin{equation}
\e^{-2t} v_{6,0} (p_1 \e^{-t},\cdots, p_6 \e^{-t}) =
\frac{1 - K((p_1+p_2+p_3) \e^{-t})}{(p_1+p_2+p_3)^2 + m^2} +
\text{5 permuations}
\end{equation}
This implies the asymptotic form
\begin{equation}
A_6 (-t; q, -q, 0,0,0,0) = (-\lambda)^2 \frac{6(1 - K(q))}{q^2}
\end{equation}
Hence, we obtain
\begin{eqnarray}
&&v_{4,1} (-t; p_1 \e^{-t},\cdots, p_4 \e^{-t}; m^2 \e^{-2t}) =
\int_0^\infty dt'\, \sum_{i=1}^4 \frac{\Delta (p_i
\e^{-(t+t')})}{p_i^2 + m^2} \, \e^{2(t+t')} v_{2,1}
(-(t+t'))\nonumber\\ &&+ \int_0^\infty dt'\, \frac{1}{2} \int_q
\Bigg[\, \frac{\Delta (q^{-(t+t')})}{q^2 + m^2} \, \e^{-2(t+t')}
v_{6,0} (q \e^{-(t+t')}, -q \e^{-(t+t')}, p_1 \e^{-(t+t')}, \cdots,
p_4 \e^{-(t+t')})\nonumber\\ &&\qquad\qquad - \frac{\Delta (q
\e^{-(t+t')})}{q^2} \cdot \frac{6(1 - K(q \e^{-(t+t')}))}{q^2} \,
\Bigg]\, - t \,\frac{1}{2} \int_q \frac{\Delta (q)}{q^2} \frac{6(1 -
K(q))}{q^2}\nonumber\\ &=& \sum_{i=1}^4 \frac{1 - K(p_i
\e^{-t})}{p_i^2 + m^2}\, \e^{2t} v_{2,1} (-t) \nonumber\\ &&\, +
\frac{1}{2} \int_q \Bigg[\: \frac{1-K(q\e^{-t})}{q^2+m^2} \left(
\frac{1 - K((p_1+p_2+q)\e^{-t})}{(p_1+p_2+q)^2+m^2} + \text{2\,
permutations}\, \right) \nonumber\\ &&\qquad\qquad\qquad - 3
\frac{(1-K(q \e^{-t}))^2}{q^4} \: \Bigg] - 3 t \int_q \frac{\Delta (q)
(1- K(q))}{q^4}
\end{eqnarray}
where we used
\begin{equation}
\Delta (q\e^{-t}) (1 - K(q\e^{-t})) = - \frac{d}{dt} \frac{1}{2} \left(1 -
K(q \e^{-t})\right)^2
\end{equation}
The expression for $v_{2,2}$ is omitted.

\section{\label{Tk}Construction of $T_k (t)$}

The $k$-th order polynomial $T_k (t)$ is defined by
\begin{equation}
\frac{d}{dt} \left( \e^{2t} T_k (t) \right) = \e^{2t} t^k
\end{equation}
By substituting
\begin{equation}
T_k (t) = \sum_{l=0}^k c_l \,t^{k-l}
\end{equation}
into the definition, we obtain a recursion relation for $c_l$ whose
solution is
\begin{eqnarray}
&&c_l = (-)^l \frac{k(k-1) \cdots (k-l+1)}{2^{l+1}} \nonumber\\
&&\quad \Longleftrightarrow\,
c_0 = \frac{1}{2},\quad c_1 = - \frac{k}{4},\quad c_2 =
\frac{k(k-1)}{8},\quad \cdots,\quad c_k = (-)^k \frac{k!}{2^{k+1}}
\end{eqnarray}

Using $T_k (t)$, we can construct a map from an $n$-th order
polynomial $P_n (t)$ to another $n$-th order polynomial:
\begin{equation}
P_n (t) = \sum_{k=0}^n P_{n,k} t^k \longrightarrow Q_n (t) =
\sum_{k=0}^n P_{n,k} T_k (t)
\end{equation}
By definition of $T_k (t)$, this has the obvious consequence
\begin{equation}
\frac{d}{dt} \left( \e^{2t} Q_n (t) \right) = \e^{2t} P_n (t)
\end{equation}
An important property of the above map is its invariance under
translation.  Namely, if the polynomial $P_n (t)$ maps to $Q_n (t)$,
then the shifted polynomial $P_n (t- \Delta t)$, where $\Delta t$ is a
constant, maps to the shifted polynomial $Q_n (t - \Delta t)$.  This
implies that the map from $P_n (t)$ to $Q_n (t)$ is defined
independent of the choice of the origin of the variable $t$.

\begin{acknowledgments}
The author would like to thank Profs.~J.~Banavar and M.~G\"unayden of
the Department of Physics at Penn State University for hosting his
visit during which the present work was completed.  This work was
partially supported by the Grant-In-Aid for Scientific Research from
the Ministry of Education, Science, and Culture, Japan (\#14340077).
\end{acknowledgments} 

\bibliography{bootstrap}

\end{document}